\documentclass{ws-ijmpa}
\usepackage{url}
\usepackage[super,compress]{cite}
\usepackage{graphicx}
\usepackage{hyperref}
\hypersetup{colorlinks,urlcolor=black,citecolor=black,linkcolor=black,filecolor=black}
\usepackage{breakurl}
\usepackage{booktabs}

\allowdisplaybreaks

\begin{document}
\markboth{Feng-Zhi Chen, Xin-Qiang Li, and Yuan-He Zou}{$\tau^- \to \omega \pi^- \nu_\tau$ decay in R$\chi$T with tensor sources}

%
\catchline{}{}{}{}{}
%

\title{$\tau^- \to \omega \pi^- \nu_\tau$ decay in R$\chi$T with tensor sources}

\author{Feng-Zhi Chen}

\address{Department of Physics, College of Physics and Optoelectronic Engineering,\\ Jinan University, Guangzhou 510632, P. R. China\\[0.1cm]
fzchen@jnu.edu.cn}

\author{Xin-Qiang Li\footnote{Speaker}, and Yuan-He Zou}

\address{Institute of Particle Physics and Key Laboratory of Quark and Lepton Physics~(MOE),\\ Central China Normal University, Wuhan, Hubei 430079, P. R. China\\[0.1cm]	
xqli@mail.ccnu.edu.cn, yuanhe99zou@mails.ccnu.edu.cn}

\maketitle

\begin{history}
\received{Day Month Year}
\revised{Day Month Year}
\end{history}

\begin{abstract}
We present a study of the $\tau^- \to \omega\pi^-\nu_\tau$ decay in the framework of low-energy effective field theory. By analyzing the $J^{PG}$ quantum numbers of the quark currents and the $\omega\pi$ final state, we find that only the Standard Model (SM) vector interaction and the non-standard tensor interaction can contribute to this decay. We construct the resonance chiral theory Lagrangian with external tensor sources and calculate both the vector and tensor form factors, with resonance couplings determined through QCD short-distance constraints, spectral function fitting, and chiral perturbation theory matching. The new physics (NP) effect is investigated in the spectral function and forward-backward asymmetry distributions. Our results show that the spectral function is dominated by the SM, while the forward-backward asymmetry, which can only arise from a non-zero tensor interaction, provides a sensitive probe of this NP effect. Future measurements at Belle II, Tera-Z, and STCF facilities are therefore strongly motivated.

\end{abstract}

\ccode{PACS numbers: 14.60.Fg, 13.35.Dx, 12.15.Ji}


\section{Introduction}	

The hadronic decay $\tau^- \to \omega\pi^-\nu_\tau$ provides an ideal laboratory for testing the so-called second-class current (SCC)~\cite{Weinberg:1958ut,Leroy:1977pq,Berger:1987ku}. The experimentally measured total branching ratio of this channel is $(1.95\pm0.06)\%$~\cite{ParticleDataGroup:2024cfk}, while the current experimental upper limit for the SCC contribution is only $1.4\times10^{-4}$ at $90\%$ confidence level (CL)~\cite{BaBar:2009jyj}. The SM estimate of the SCC-induced branching ratio via isospin breaking is $(1.5\sim2.8)\times10^{-5}$~\cite{Paver:2012tq}, leaving substantial room for genuine SCC contributions from non-standard interactions. However, as can be seen from these numbers, even if genuine SCC contributions exist in this decay, the upper limit is two orders of magnitude smaller than that of the total branching ratio. Thus, we can safely neglect the SCC contributions for the moment and assume that the $\tau^-\to\omega\pi^-\nu_\tau$ decay is dominated by the first-class current (FCC) non-standard interactions, while the $G$-parity is assumed to be strictly conserved.

To study possible NP effects in a model-independent way, we employ the framework of low-energy effective field theory (LEFT)~\cite{Jenkins:2017jig,Jenkins:2017dyc}, which is invariant under Lorentz transformations and respects the $SU(3)_C \otimes U(1)_{em}$ gauge symmetry. Assuming $G$-parity conservation (which assures that only the FCC contributes) and analyzing the $J^{PG}$ quantum numbers of the quark currents and the $\omega\pi$ system, we can show that, besides the SM vector contribution, only a non-standard tensor interaction can have a non-zero impact on the $\tau^- \to \omega\pi^-\nu_\tau$ decay.

To estimate the tensor contribution, we construct the resonance chiral theory (R$\chi$T) Lagrangian with external tensor sources and calculate the $\omega\pi$ tensor form factors. The resonance couplings are determined by combining QCD short-distance constraints, fits to the spectral function, and matching between $\mathcal{O}(p^4)$ R$\chi$T operators and $\mathcal{O}(p^6)$ chiral perturbation theory ($\chi$PT) operators.

\section{Theoretical Framework}

\subsection{LEFT Lagrangian for $\tau^- \to \bar{u}d\nu_\tau$}

The $\tau^- \to \omega\pi^-\nu_\tau$ decay proceeds through the quark-level $\tau^- \to \bar{u}d\nu_\tau$ transition. The most generic LEFT Lagrangian describing this transition is given by~\cite{Garces:2017jpz,Cirigliano:2017tqn,Miranda:2018cpf,Cirigliano:2018dyk,Rendon:2019awg,Chen:2019vbr,Gonzalez-Solis:2019lze,Gonzalez-Solis:2020jlh,Chen:2020uxi,Arroyo-Urena:2021dfe,Arroyo-Urena:2021nil,Chen:2021udz,Cirigliano:2021yto}
\begin{align}
\label{eq:LEFT}
\mathcal{L}_\mathrm{eff} = &-\frac{G_F^0 V_{ud}}{\sqrt{2}} \left(1+\epsilon_L +\epsilon_R\right) \Big\{ \bar{\tau} \gamma_\mu (1-\gamma_5) \nu_\tau \cdot \left[ \bar{u} \gamma^\mu d -(1-2\hat{\epsilon}_R) \bar{u} \gamma^\mu \gamma_5 d \right] \nonumber\\
& + \bar{\tau} (1-\gamma_5) \nu_\tau \cdot \left[ \hat{\epsilon}_S \bar{u} d -\hat{\epsilon}_P \bar{u} \gamma_5 d \right] + 2\hat{\epsilon}_T \bar{\tau} \sigma_{\mu\nu} (1-\gamma_5) \nu_\tau \cdot \bar{u} \sigma^{\mu\nu} d \Big\} + \mathrm{h.c.}\,,
\end{align}
where $G_F^0$ is the Fermi constant, $V_{ud}$ denotes the CKM matrix element, and $\hat{\epsilon}_i=\epsilon_i/(1+\epsilon_L +\epsilon_R)$~($i=R,S,P,T$) parameterize the non-standard contributions normalized to the SM vector one. 

\subsection{$J^{PG}$ Analysis}

Among all the quark-current operators in Eq.~\eqref{eq:LEFT}, only those with the same $J^{PG}$ quantum numbers as the $\omega\pi^-$ final state contribute to the $\tau^- \to \omega\pi^-\nu_\tau$ decay. We summarize in Table~\ref{tab:JPG} the $J^{PG}$ quantum numbers of them and the $\omega\pi^-$ system with three different orbital angular momenta, from which one can see that only the vector quark-current operator $\mathcal{O}_V=\bar{u} \gamma^\mu d$ with $J^{PG}=1^{-+}$ and the tensor operator $\mathcal{O}_T=\bar{\tau} \sigma_{\mu\nu} (1-\gamma_5) \nu_\tau \cdot \bar{u} \sigma^{\mu\nu} d$ with $J^{PG}=1^{-+}$ contribute to the decay~\cite{Chen:2024xqt}. 

\begin{table}[htbp]
\tabcolsep 0.21in
	\renewcommand{\arraystretch}{1.3}
	\tbl{\small $J^{PG}$ quantum numbers of the quark-current operators and the $\omega\pi$ system.\label{tab:JPG}}
	{\setlength{\tabcolsep}{4pt} 
	\resizebox{\textwidth}{!}{
  \begin{tabular}{|c|c|c|c|c|c|c|c|c|}
  \toprule
   & $\mathcal{O}_S$ & $\mathcal{O}_P$ & $\mathcal{O}_V$ & $\mathcal{O}_A$ & $\mathcal{O}_T$ & $\omega\pi\mid_{L=0}$& $\omega\pi\mid_{L=1}$& $\omega\pi\mid_{L=2}$\\
  \midrule
  $J^{PG}$ &  $0^{+-}$ & $0^{--}$ & $0^{++}\oplus 1^{-+}$ & $0^{--}\oplus 1^{+-}$ & $1^{++}\oplus 1^{-+}$& $1^{++}$ & $0^{-+}\oplus 1^{-+}\oplus 2^{-+}$ & $1^{++}\oplus2^{++}\oplus3^{++}$\\ 
  \bottomrule
  \end{tabular} }}
\end{table}

It is worth emphasizing that, unlike $\tau^- \to \omega\pi^-\nu_\tau$, a general $\tau\to VP\nu_\tau$ decay, with $V=\rho, K^\ast, \omega, \phi$ and $P=\pi, K, \eta^{(\prime)}$, is not constrained by $G$ parity. In this case, only the $J^P$ quantum numbers in Table~\ref{tab:JPG} are relevant. Consequently, $\mathcal{O}_S=\bar{u} d$ does not contribute to any $VP$ final state. $\mathcal{O}_P=\bar{u} \gamma_5 d$ and $\mathcal{O}_V$ provide nonzero contributions to the $J^P=0^-$ and $J^P=1^-$ $P$ waves, respectively; $\mathcal{O}_A=\bar{u} \gamma^\mu \gamma_5 d$ contributes to $J^P=1^+$ $S$ and $D$, as well as $J^P=0^-$ $P$ waves, while $\mathcal{O}_T$ to $J^P=1^+$ $S$ and $D$, as well as $J^P=1^-$ $P$ waves.

\subsection{Observables}

The doubly differential decay width of $\tau^-(p) \to \omega(p_2) \pi^-(p_1) \nu_\tau(p_3)$ is given by
\begin{align}\label{eq:rate}
\frac{d^2\Gamma(\tau^- \to \omega \pi^- \nu_\tau)}{ds \, d\cos\theta}=A(s)+B(s)\cos\theta +C(s)\cos^2\theta\,,
\end{align}
where $\theta$ is the angle between the directions of $\omega$ and $\tau$ in the $\omega\pi$ rest frame, and
\begin{align} 
\label{eq:A} 
A(s) =\,& \frac{G_F^2 S_{EW} |V_{ud}|^2}{2048 \pi^3 s^2 m_\tau^3} (s-m_\tau^2)^2 \lambda^{\frac{3}{2}}(s, M_\omega^2, M_\pi^2) \notag \\[1mm]
     & \times \Big\lbrace  \left( m_\tau^2 +s \right) |F_{V}(s)|^2 + 16 m_\tau \hat{\epsilon}_T \mathrm{Re} \left[ F_{V}(s) \left( F_{T3}^\ast(s)-F_{T2}^\ast(s) \right) \right] \Big\rbrace\,,\\[2mm]
\label{eq:B} 
B(s) =\,& -\frac{G_F^2 S_{EW} |V_{ud}|^2 }{128 \pi^3 s^2 m_\tau^2} (s-m_\tau^2)^2 \lambda(s, M_\omega^2, M_\pi^2) \notag \\[1mm]
     & \times \hat{\epsilon}_T \Big\lbrace \left( s-\Delta_{\omega \pi} \right) \mathrm{Re}\left[ F_{V}(s) F_{T2}^\ast(s)\right] + \left( s+\Delta_{\omega \pi} \right) \mathrm{Re}\left[ F_{V}(s) F_{T3}^\ast(s)\right] \Big\rbrace\,,  \\[2mm]  
\label{eq:C} 
C(s) =\,& \frac{G_F^2 S_{EW} |V_{ud}|^2}{2048 \pi^3 s^2 m_\tau^3} (s-m_\tau^2)^3 \lambda^{\frac{3}{2}}(s, M_\omega^2, M_\pi^2) |F_{V}(s)|^2\,.   
\end{align}
Here $F_V(s)$ and $F_{Ti}(s)$ ($i=1,2,3$) are the vector and tensor form factors, respectively. From Eq.~\eqref{eq:rate}, one can define the spectral function
\begin{align}
v(s) = \frac{32 \pi^2 m_\tau^3}{G_F^{2} |V_{ud}|^2 (m_\tau^2 - s)^2 (m_\tau^2 + 2s)} \frac{d\Gamma(\tau^- \to \omega \pi^- \nu_\tau)}{ds}\,,
\end{align}
as well as the forward-backward asymmetry
\begin{align}
A_\mathrm{FB}(s) = \frac{3B(s)}{6A(s)+2C(s)}\,.
\end{align}
In the SM, $A_\mathrm{FB}(s)$ vanishes, and a non-zero value arises only with a non-zero $\hat{\epsilon}_T$. As this observable is usually measured in bins of the $\omega \pi$ invariant mass, one can also define a bin-dependent forward-backward asymmetry
\begin{align}
\langle A_{\mathrm{FB}}\rangle_i =
\frac{\int_{s_1^i}^{s_2^i} B(s)ds}
{\int_{s_1^i}^{s_2^i} \left[ 2A(s) + \frac{2}{3}C(s) \right] ds}\,,
\label{eq:FBAbin}
\end{align}
where $s_1^i$ and $s_2^i$ denote the lower and upper boundaries of the $i$-th bin, respectively.

\section{$\omega\pi$ Form Factors in R$\chi$T}

The $\omega\pi$ vector form factor $F_V(s)$ has been calculated in the R$\chi$T framework in Ref.~\cite{Guo:2008sh}. To calculate the tensor form factors $F_{Ti}(s)$ in the same framework, we have to construct the R$\chi$T Lagrangian with external tensor sources.

The relevant R$\chi$T Lagrangian at leading order in the large-$N_C$ expansion reads~\cite{Weinberg:1978kz,Gasser:1983yg,Gasser:1984gg,Ecker:1988te,Ecker:1989yg,Ruiz-Femenia:2003jdx,Cirigliano:2004ue}
\begin{align}
\mathcal{L}=\mathcal{L}_\chi^{(2)} + \mathcal{L}_{kin}(V) + \mathcal{L}_{2V} + \mathcal{L}_{VVP}+ \mathcal{L}_{VJP} + \mathcal{L}_{VT} + \mathcal{L}_{VTP}\,.
\end{align}
The new terms involving external tensor sources are given by~\cite{Chen:2024xqt}
\begin{align}
	\mathcal{L}_{VT} =\,& F^T_V \left\langle  V_{\mu\nu} t^{\mu\nu}_+ \right\rangle\,, \label{eq:VT} \\[0.2cm]
	\mathcal{L}_{V T P} =\, & b_1 \epsilon_{\mu \nu \rho \sigma}\left\langle\left\{V^{\mu \nu}, t_{+}^{\rho \alpha}\right\} \nabla_\alpha u^\sigma\right\rangle +b_2 \epsilon_{\mu \nu \rho \sigma}\langle \{V^{\mu \alpha},t_{+}^{\nu\rho}\} \nabla_\alpha u^\sigma\rangle  \nonumber \\[1mm]
	&+i b_3 \epsilon_{\mu \nu \rho \sigma}\left\langle\left\{V^{\mu \nu}, t_{+}^{\rho \sigma}\right\} \chi_{-}\right\rangle +i b_4 \epsilon_{\mu \nu \rho \sigma}\left\langle\left\{V^{\mu \nu}, t_{-}^{\rho \sigma}\right\} \chi_{+}\right\rangle \nonumber \\[1mm]
	& +b_{5} \epsilon_{\mu \nu \rho \sigma}\left\langle\left\{\nabla_\alpha V^{\mu \nu}, t_{+}^{\rho \alpha}\right\} u^\sigma\right\rangle +b_{6} \epsilon_{\mu \nu \rho \sigma}\left\langle\left\{\nabla_\alpha V^{\mu \alpha}, t_{+}^{\nu\rho}\right\} u^\sigma\right\rangle \nonumber \\[1mm]
	&+b_{7} \epsilon_{\mu \nu \rho \sigma}\left\langle\left\{\nabla^\mu V^{\nu \rho}, t_{+}^{\sigma \alpha}\right\} u_\alpha\right\rangle +i b_8 \epsilon_{\mu \nu \rho \sigma}\left\langle V^{\mu \nu}t_{-}^{\rho \sigma}\right\rangle\langle \chi_{+} \rangle \nonumber \\[1mm]
	&+b_9\epsilon_{\mu \nu \rho \sigma}\langle V^{\mu \nu} \nabla_\alpha u^\rho\rangle\langle t_{+}^{\sigma \alpha} \rangle +b_{10} \epsilon_{\mu \nu \rho \sigma}\langle V^{\mu\nu}\nabla^{\rho}u_\alpha \rangle\langle  t_{+}^{\sigma\alpha} \rangle \nonumber \\[1mm]
	&+i b_{11} \epsilon_{\mu \nu \rho \sigma}\left\langle V^{\mu \nu}  \chi_{-}\right\rangle\langle t_{+}^{\rho \sigma}\rangle +i b_{12} \epsilon_{\mu \nu \rho \sigma}\left\langle V^{\mu \nu}\chi_{+}\right\rangle\langle t_{-}^{\rho \sigma} \rangle \nonumber \\[1mm]
	& + b_{13} \epsilon_{\mu \nu \rho \sigma}\left\langle\nabla_\alpha V^{\mu \nu} u^\rho\right\rangle\langle t_{+}^{\sigma \alpha} \rangle +b_{14} \epsilon_{\mu \nu \rho \sigma}\left\langle\nabla_\alpha V^{\mu \alpha}u^\nu\right\rangle\langle t_{+}^{\rho \sigma}\rangle \nonumber\\[1mm]
	& +b_{15} \epsilon_{\mu \nu \rho \sigma}g_{\alpha\beta}\left\langle\nabla^\mu V^{ \nu\alpha} u^\rho\right\rangle\langle t_{+}^{\sigma \beta} \rangle\,, \label{eq:VTP}
\end{align}
where $F_V^T$ and $b_i$ ($i=1,\cdots,15$) represent the corresponding resonance couplings, and $t_{\pm}^{\mu\nu} = u^\dagger t^{\mu\nu} u^\dagger \pm u\, {t^{\mu\nu}}^\dagger u$~\cite{Cata:2007ns}, with $t^{\mu\nu}$ and $t^{\mu\nu\dagger}$ being the external tensor sources. 

With Eqs.~\eqref{eq:VT} and \eqref{eq:VTP} at hand, one can calculate the $\omega\pi$ vector and tensor form factors. After imposing the QCD short-distance constraints, which require the form factors to vanish smoothly as $s \to \infty$, we arrive at the following expressions:
\begin{align}
	\label{eq:V(s)} F_{V}(s)=& \frac{2 \sqrt{2}}{F M_\omega} \left( 2 d_3 F_V +d_s F_{V_1} \right) +\frac{4\sqrt{2} F_V}{F M_\omega} \left[ d_{12} M_\pi^2 + d_3 (s+\Delta_{\omega\pi}) \right] D_\rho(s) \nonumber \\[1mm]
	&+\frac{2\sqrt{2} F_{V_1}}{F M_\omega} \left[d_m M_\pi^2 + d_M M_\omega^2+d_s s\right] D_{\rho^\prime}(s)\,, \\[2mm]
	\label{eq:T_1(s)} F_{T1}(s)=& \frac{4 F^T_V}{F M_\omega M_\rho^2} \left[ d_{12} M_\pi^2 + d_3 \left( \Delta_{\omega\pi} +M_\rho^2 \right) \right] D_\rho(s)  \nonumber \\[1mm]
	&-\frac{4 F^T_{V_1}}{F M_\omega M_{\rho^\prime}^2} \left[ d_d M_\omega^2 -\left( d_b +4 d_f \right) M_\pi^2 -\left( d_a-d_b \right) M_{\rho^\prime}^2 \right] D_{\rho^\prime}(s)\,,  \\[2mm]
	\label{eq:T_2(s)} F_{T2}(s)=& -\frac{2 F^T_V}{F M_\omega M_\rho^2}  \left\lbrace \left[ d_{12} M_\pi^2 -d_3 ( \Sigma_{\omega\pi} -M_\rho^2 ) \right] + \left[ d_{12} M_\pi^2 (s+\Delta_{\omega\pi}) \right. \right. \nonumber\\[1mm]
	&\left. \left. -d_3 ( s\Sigma_{\omega\pi}-\Delta_{\omega\pi}^2 + M_\rho^2 ( 2 M_\omega^2 +s+\Delta_{\omega\pi} ) ) \right] D_\rho(s) \right\rbrace \nonumber\\[1mm]	
	& -\frac{2 F^T_{V_1}}{F M_\omega M_{\rho^\prime}^2} \left\lbrace \left[ d_d M_\omega^2 +( d_b +4 d_f ) M_\pi^2 +( d_a-d_b ) M_{\rho^\prime}^2 \right] \right. \nonumber \\[1mm]
	&\times ( 1+(s-\Sigma_{\omega\pi}) D_{\rho^\prime}(s) ) + 2 M_\omega^2 [ ( d_b+d_d+4 d_f ) M_\pi^2 \nonumber \\[1mm]
	&\left. +( d_a-d_b-d_d ) M_{\rho^\prime}^2 ] D_{\rho^\prime}(s) \right\rbrace\,, \\[2mm]
	\label{eq:T_3(s)} F_{T3}(s)=& -\frac{2 F^T_V}{F M_\omega M_\rho^2} \left\lbrace \left[ d_{12} M_\pi^2 -d_3 ( \Sigma_{\omega\pi} +M_\rho^2 ) \right] + \left[ d_{12} M_\pi^2 ( s+\Delta_{\omega\pi} \right. \right. \nonumber\\[1mm]
	&\left. \left. -2 M_\rho^2 ) -d_3 ( s\Sigma_{\omega\pi}-\Delta_{\omega\pi}^2 + M_\rho^2(s-\Sigma_{\omega\pi}) ) \right] D_{\rho}(s) \right\rbrace  \nonumber \\[1mm]
	&-\frac{2 F^T_{V_1}}{F M_\omega M_{\rho^\prime}^2} \left\lbrace \left[ d_d M_\omega^2 +( d_b +4 d_f ) M_\pi^2 -( d_a-d_b ) M_{\rho^\prime}^2 \right] \right. \nonumber \\[1mm]
	&\times ( 1+(s-\Sigma_{\omega\pi}) D_{\rho^\prime}(s) ) + 2 M_\pi^2 [ ( d_b+d_d+4 d_f ) M_\omega^2 \nonumber \\[1mm]
	&\left. -( d_a+4 d_f ) M_{\rho^\prime}^2 ] D_{\rho^\prime}(s) \right\rbrace\,,
\end{align}
where $D_V(s)$ denotes the propagator of the intermediate vector resonance $V$ with an energy-dependent decay width~\cite{Guo:2008sh}.

The resonance couplings $d_i$ are determined by fitting the SM-predicted spectral function to the experimental data from CLEO~\cite{CLEO:1999heg}, yielding
\begin{equation}
d_3 =-0.229 \pm 0.008\,,\; d_s =-0.259 \pm 0.039\,,\; d_m M_\pi^2+d_M M_\omega^2=0.525\pm0.067\,,
\end{equation}
with $\chi^2_\mathrm{min}/\mathrm{d.o.f.}\simeq 2.0$. The remaining resonance couplings are constrained by matching the $\mathcal{O}(p^4)$ odd-intrinsic-parity R$\chi$T Lagrangian to the $\mathcal{O}(p^6)$ $\chi$PT sector~\cite{Bijnens:2001bb}. Detailed derivations of these constraints can be found in Ref.~\cite{Chen:2024xqt}, and the resulting numerical values for all these parameters are summarized in Table~\ref{tab:couplings}.

\begin{table}[thbp]
\centering
\tbl{Numerical results of the resonance couplings. \label{tab:couplings}}
{\setlength{\tabcolsep}{20pt} 
\renewcommand\arraystretch{1.3} 
\small 
\begin{tabular}{|l|l|l|}
\toprule
   $c_1 = -7.44\times10^{-3}$ & $c_7 = -3.40\times10^{-1}$ & $d_c = -1.29\times10^{-1}$ \\
   $c_2 = -3.45\times10^{-3}$ & $d_1 = 6.08\times10^{-1}$  & $d_d = -4.33\times10^{-1}$ \\
   $c_3 = 1.86\times10^{-3}$  & $d_2 = -7.99\times10^{-2}$ & $d_e = 7.19\times10^{-1}$  \\
   $c_4 = -9.59\times10^{-3}$ & $d_3 = -2.29\times10^{-1}$ & $d_f = 7.49\times10^{-1}$  \\
   $c_5 = 3.99\times10^{-3}$  & $d_4 = 0$                  & $d_a = 2.73$               \\
   $c_6 = 2.16\times10^{-2}$  & $d_b = 2.86$               &                            \\
\bottomrule
\end{tabular}}
\end{table}

\section{Numerical Results and Discussion}

With the form factors at hand, we can set constraints on the tensor coefficient $\hat{\epsilon}_T$ by using the measured branching ratio $\mathcal{B}(\tau^- \to \omega \pi^- \nu_\tau)_\mathrm{exp}=(1.95\pm0.06)\%$~\cite{ParticleDataGroup:2024cfk} and the spectral function data in 16 bins from CLEO~\cite{CLEO:1999heg}. Our fitting result reads
\begin{align} \label{eq:epsilon_T_bound}
\hat{\epsilon}_T=(0.3\pm4.9)\times10^{-3}\,,
\end{align}
with $\chi^2_\mathrm{min}/\mathrm{d.o.f.}\simeq1.7$. It is observed that this constraint is about one order of magnitude stronger than the previous bounds obtained from a simultaneous fit to other exclusive hadronic $\tau$ decays~\cite{Gonzalez-Solis:2020jlh}.

\begin{figure}[tbhp]
\centering
\includegraphics[width=0.48\textwidth]{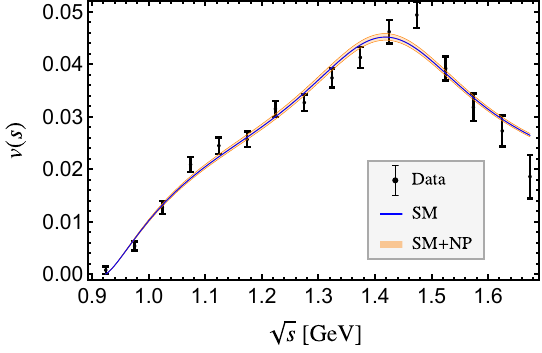}\;
\includegraphics[width=0.485\textwidth]{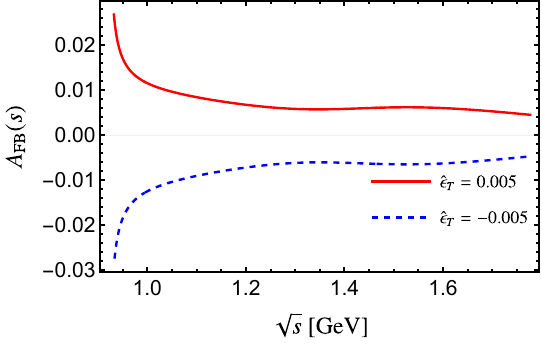}
\caption{Left: the $\sqrt{s}$ distribution of the spectral function $v(s)$ with (orange band) and without (blue line) the NP contribution, where the bin data is taken from Ref.~\cite{CLEO:1999heg}. Right: the $\sqrt{s}$ distribution of the forward-backward asymmetry $A_\mathrm{FB}(s)$ predicted with two different values of the tensor coefficient, $\hat{\epsilon}_T=0.005$ (red solid) and $\hat{\epsilon}_T=-0.005$ (blue dashed). \label{fig:NPimpact}.}
\label{fig:NPimpact}
\end{figure}

We show in Fig.~\ref{fig:NPimpact} the NP impacts on the spectral function $v(s)$ and the forward-backward asymmetry $A_\mathrm{FB}(s)$. The left panel demonstrates that the NP contribution to the spectral function is negligible compared to the SM prediction. However, as can be seen from the right panel, the forward-backward asymmetry $A_\mathrm{FB}(s)$ only arises from a non-zero tensor interaction, making it an ideal observable to probe the non-standard tensor effect. 

Finally, we provide a rough estimate of the sensitivity to non-standard tensor interaction at the future $\tau$ facilities, such as the STCF~\cite{Achasov:2023gey}. Operating at a center-of-mass energy of $\sqrt{s}=4.009~\textit{GeV}$ with an integrated annual luminosity of $1~\text{ab}^{-1}$, the STCF is expected to produce $3.5 \times 10^9$ $\tau^+\tau^-$ pairs per year. Assuming a total signal reconstruction efficiency of $\mathcal{O}(10\%)$, one expects to collect approximately $N_{\text{sig}} \sim 6.8 \times 10^6$ signal events of the $\tau^- \to \omega \pi^- \nu_\tau$ decay annually. On the other hand, integrating Eq.~\eqref{eq:FBAbin} over the full $\omega\pi$ invariant-mass range, we obtain an integrated forward-backward asymmetry
\begin{equation}
       \langle A_{\mathrm{FB}}\rangle \simeq 1.4\,\hat{\epsilon}_T\,.
\end{equation} 
Assuming the measurement is limited primarily by the statistical uncertainty, $\delta A_{\mathrm{FB}} \approx 1/\sqrt{N_{\text{sig}}}$, the expected sensitivity to $\hat{\epsilon}_T$ could reach $\mathcal{O}(10^{-4})$. This represents an improvement over the current experimental bound given by Eq.~\eqref{eq:epsilon_T_bound} by more than an order of magnitude.

\section{Conclusion}

In this work, we have performed a comprehensive study of the NP effects in the $\tau^- \to \omega\pi^-\nu_\tau$ decay within the LEFT framework. Based on the $J^{PG}$ quantum number analysis, we demonstrate that beyond the SM vector interaction, only the non-standard tensor interaction contributes to this decay. To evaluate the corresponding hadronic matrix elements, we have constructed for the first time the R$\chi$T Lagrangian $\mathcal{L}_{VTP}$ with external tensor sources, enabling a consistent calculation of the $\omega\pi$ tensor form factors. Our analysis yields the most stringent constraint to date on the tensor coupling, $\hat{\epsilon}_T = (0.3 \pm 4.9) \times 10^{-3}$, from exclusive hadronic $\tau$ decays. While the spectral function remains dominated by the SM, we show that the forward-backward asymmetry $A_\mathrm{FB}(s)$ serves as a powerful null-test probe of NP, vanishing identically in the SM but exhibiting a high sensitivity to the non-standard tensor interaction. Consequently, precise measurements of the observable $A_\mathrm{FB}(s)$ at the future facilities, such as Belle II~\cite{Belle-II:2018jsg}, Tera-Z~\cite{CEPCStudyGroup:2018ghi,FCC:2018byv}, and STCF~\cite{Achasov:2023gey}, are strongly motivated to probe such a non-zero tensor interaction.

\section*{Acknowledgements}

This work is supported by the National Natural Science Foundation of China under Grant Nos.~12475094 and 12135006. FC is also supported by the Fundamental Research Funds for the Central Universities (11623330) and the 2024 Guangzhou Basic and Applied Basic Research Scheme Project (2024A04J4190).

\bibliographystyle{ws-ijmpa}
\bibliography{reference}

\begin{thebibliography}{10}
\expandafter\ifx\csname urlstyle\endcsname\relax
  \providecommand{\doi}[1]{doi:\discretionary{}{}{}#1}\else
  \providecommand{\doi}{doi:\discretionary{}{}{}\begingroup \urlstyle{rm}\Url}\fi

\bibitem{Weinberg:1958ut}
S.~Weinberg, {\em Phys. Rev.} {\bf 112}, 1375  (1958), \doi{10.1103/PhysRev.112.1375}.

\bibitem{Leroy:1977pq}
C.~Leroy and J.~Pestieau, {\em Phys. Lett. B} {\bf 72}, 398  (1978), \doi{10.1016/0370-2693(78)90148-X}.

\bibitem{Berger:1987ku}
E.~L. Berger and H.~J. Lipkin, {\em Phys. Rev. Lett.} {\bf 59},   1394  (1987), \doi{10.1103/PhysRevLett.59.1394}.

\bibitem{ParticleDataGroup:2024cfk}
 Particle Data Group Collaboration (S.~Navas {\em et~al.}), {\em Phys. Rev. D} {\bf 110},   030001  (2024), \doi{10.1103/PhysRevD.110.030001}.

\bibitem{BaBar:2009jyj}
 BaBar Collaboration (B.~Aubert {\em et~al.}), {\em Phys. Rev. Lett.} {\bf 103},   041802  (2009), \href{http://arxiv.org/abs/0904.3080}{{\ttfamily arXiv:0904.3080 [hep-ex]}}, \doi{10.1016/j.nuclphysbps.2009.03.021}.

\bibitem{Paver:2012tq}
N.~Paver and Riazuddin, {\em Phys. Rev. D} {\bf 86},   037302  (2012), \href{http://arxiv.org/abs/1205.6636}{{\ttfamily arXiv:1205.6636 [hep-ph]}}, \doi{10.1103/PhysRevD.86.037302}.

\bibitem{Jenkins:2017jig}
E.~E. Jenkins, A.~V. Manohar and P.~Stoffer, {\em JHEP} {\bf 03},   016  (2018), \href{http://arxiv.org/abs/1709.04486}{{\ttfamily arXiv:1709.04486 [hep-ph]}}, \doi{10.1007/JHEP03(2018)016}, [Erratum: JHEP 12, 043 (2023)].

\bibitem{Jenkins:2017dyc}
E.~E. Jenkins, A.~V. Manohar and P.~Stoffer, {\em JHEP} {\bf 01},   084  (2018), \href{http://arxiv.org/abs/1711.05270}{{\ttfamily arXiv:1711.05270 [hep-ph]}}, \doi{10.1007/JHEP01(2018)084}, [Erratum: JHEP 12, 042 (2023)].

\bibitem{Garces:2017jpz}
E.~A. Garc\'es, M.~Hern\'andez~Villanueva, G.~L\'opez~Castro and P.~Roig, {\em JHEP} {\bf 12},   027  (2017), \href{http://arxiv.org/abs/1708.07802}{{\ttfamily arXiv:1708.07802 [hep-ph]}}, \doi{10.1007/JHEP12(2017)027}.

\bibitem{Cirigliano:2017tqn}
V.~Cirigliano, A.~Crivellin and M.~Hoferichter, {\em Phys. Rev. Lett.} {\bf 120},   141803  (2018), \href{http://arxiv.org/abs/1712.06595}{{\ttfamily arXiv:1712.06595 [hep-ph]}}, \doi{10.1103/PhysRevLett.120.141803}.

\bibitem{Miranda:2018cpf}
J.~A. Miranda and P.~Roig, {\em JHEP} {\bf 11},   038  (2018), \href{http://arxiv.org/abs/1806.09547}{{\ttfamily arXiv:1806.09547 [hep-ph]}}, \doi{10.1007/JHEP11(2018)038}.

\bibitem{Cirigliano:2018dyk}
V.~Cirigliano, A.~Falkowski, M.~Gonz\'alez-Alonso and A.~Rodr\'\i{}guez-S\'anchez, {\em Phys. Rev. Lett.} {\bf 122},   221801  (2019), \href{http://arxiv.org/abs/1809.01161}{{\ttfamily arXiv:1809.01161 [hep-ph]}}, \doi{10.1103/PhysRevLett.122.221801}.

\bibitem{Rendon:2019awg}
J.~Rend\'on, P.~Roig and G.~Toledo~S\'anchez, {\em Phys. Rev. D} {\bf 99},   093005  (2019), \href{http://arxiv.org/abs/1902.08143}{{\ttfamily arXiv:1902.08143 [hep-ph]}}, \doi{10.1103/PhysRevD.99.093005}.

\bibitem{Chen:2019vbr}
F.-Z. Chen, X.-Q. Li, Y.-D. Yang and X.~Zhang, {\em Phys. Rev. D} {\bf 100},   113006  (2019), \href{http://arxiv.org/abs/1909.05543}{{\ttfamily arXiv:1909.05543 [hep-ph]}}, \doi{10.1103/PhysRevD.100.113006}.

\bibitem{Gonzalez-Solis:2019lze}
S.~Gonz\`alez-Sol\'\i{}s, A.~Miranda, J.~Rend\'on and P.~Roig, {\em Phys. Rev. D} {\bf 101},   034010  (2020), \href{http://arxiv.org/abs/1911.08341}{{\ttfamily arXiv:1911.08341 [hep-ph]}}, \doi{10.1103/PhysRevD.101.034010}.

\bibitem{Gonzalez-Solis:2020jlh}
S.~Gonz\`alez-Sol\'\i{}s, A.~Miranda, J.~Rend\'on and P.~Roig, {\em Phys. Lett. B} {\bf 804},   135371  (2020), \href{http://arxiv.org/abs/1912.08725}{{\ttfamily arXiv:1912.08725 [hep-ph]}}, \doi{10.1016/j.physletb.2020.135371}.

\bibitem{Chen:2020uxi}
F.-Z. Chen, X.-Q. Li and Y.-D. Yang, {\em JHEP} {\bf 05},   151  (2020), \href{http://arxiv.org/abs/2003.05735}{{\ttfamily arXiv:2003.05735 [hep-ph]}}, \doi{10.1007/JHEP05(2020)151}.

\bibitem{Arroyo-Urena:2021dfe}
M.~A. Arroyo-Ure\~na, G.~Hern\'andez-Tom\'e, G.~L\'opez-Castro, P.~Roig and I.~Rosell, {\em JHEP} {\bf 02},   173  (2022), \href{http://arxiv.org/abs/2112.01859}{{\ttfamily arXiv:2112.01859 [hep-ph]}}, \doi{10.1007/JHEP02(2022)173}.

\bibitem{Arroyo-Urena:2021nil}
M.~A. Arroyo-Ure\~na, G.~Hern\'andez-Tom\'e, G.~L\'opez-Castro, P.~Roig and I.~Rosell, {\em Phys. Rev. D} {\bf 104},   L091502  (2021), \href{http://arxiv.org/abs/2107.04603}{{\ttfamily arXiv:2107.04603 [hep-ph]}}, \doi{10.1103/PhysRevD.104.L091502}.

\bibitem{Chen:2021udz}
F.-Z. Chen, X.-Q. Li, S.-C. Peng, Y.-D. Yang and H.-H. Zhang, {\em JHEP} {\bf 01},   108  (2022), \href{http://arxiv.org/abs/2107.12310}{{\ttfamily arXiv:2107.12310 [hep-ph]}}, \doi{10.1007/JHEP01(2022)108}.

\bibitem{Cirigliano:2021yto}
V.~Cirigliano, D.~D\'\i{}az-Calder\'on, A.~Falkowski, M.~Gonz\'alez-Alonso and A.~Rodr\'\i{}guez-S\'anchez, {\em JHEP} {\bf 04},   152  (2022), \href{http://arxiv.org/abs/2112.02087}{{\ttfamily arXiv:2112.02087 [hep-ph]}}, \doi{10.1007/JHEP04(2022)152}.

\bibitem{Chen:2024xqt}
F.-Z. Chen, X.-Q. Li, S.-C. Peng, Y.-D. Yang and Y.-H. Zou, {\em JHEP} {\bf 08},   201  (2024), \href{http://arxiv.org/abs/2407.00700}{{\ttfamily arXiv:2407.00700 [hep-ph]}}, \doi{10.1007/JHEP08(2024)201}.

\bibitem{Guo:2008sh}
Z.-H. Guo, {\em Phys. Rev. D} {\bf 78},   033004  (2008), \href{http://arxiv.org/abs/0806.4322}{{\ttfamily arXiv:0806.4322 [hep-ph]}}, \doi{10.1103/PhysRevD.78.033004}.

\bibitem{Weinberg:1978kz}
S.~Weinberg, {\em Physica A} {\bf 96}, 327  (1979), \doi{10.1016/0378-4371(79)90223-1}.

\bibitem{Gasser:1983yg}
J.~Gasser and H.~Leutwyler, {\em Annals Phys.} {\bf 158},   142  (1984), \doi{10.1016/0003-4916(84)90242-2}.

\bibitem{Gasser:1984gg}
J.~Gasser and H.~Leutwyler, {\em Nucl. Phys. B} {\bf 250}, 465  (1985), \doi{10.1016/0550-3213(85)90492-4}.

\bibitem{Ecker:1988te}
G.~Ecker, J.~Gasser, A.~Pich and E.~de~Rafael, {\em Nucl. Phys. B} {\bf 321}, 311  (1989), \doi{10.1016/0550-3213(89)90346-5}.

\bibitem{Ecker:1989yg}
G.~Ecker, J.~Gasser, H.~Leutwyler, A.~Pich and E.~de~Rafael, {\em Phys. Lett. B} {\bf 223}, 425  (1989), \doi{10.1016/0370-2693(89)91627-4}.

\bibitem{Ruiz-Femenia:2003jdx}
P.~D. Ruiz-Femenia, A.~Pich and J.~Portoles, {\em JHEP} {\bf 07},   003  (2003), \href{http://arxiv.org/abs/hep-ph/0306157}{{\ttfamily arXiv:hep-ph/0306157}}, \doi{10.1088/1126-6708/2003/07/003}.

\bibitem{Cirigliano:2004ue}
V.~Cirigliano, G.~Ecker, M.~Eidemuller, A.~Pich and J.~Portoles, {\em Phys. Lett. B} {\bf 596}, 96  (2004), \href{http://arxiv.org/abs/hep-ph/0404004}{{\ttfamily arXiv:hep-ph/0404004}}, \doi{10.1016/j.physletb.2004.06.082}.

\bibitem{Cata:2007ns}
O.~Cata and V.~Mateu, {\em JHEP} {\bf 09},   078  (2007), \href{http://arxiv.org/abs/0705.2948}{{\ttfamily arXiv:0705.2948 [hep-ph]}}, \doi{10.1088/1126-6708/2007/09/078}.

\bibitem{CLEO:1999heg}
 CLEO Collaboration (K.~W. Edwards {\em et~al.}), {\em Phys. Rev. D} {\bf 61},   072003  (2000), \href{http://arxiv.org/abs/hep-ex/9908024}{{\ttfamily arXiv:hep-ex/9908024}}, \doi{10.1103/PhysRevD.61.072003}.

\bibitem{Bijnens:2001bb}
J.~Bijnens, L.~Girlanda and P.~Talavera, {\em Eur. Phys. J. C} {\bf 23}, 539  (2002), \href{http://arxiv.org/abs/hep-ph/0110400}{{\ttfamily arXiv:hep-ph/0110400}}, \doi{10.1007/s100520100887}.

\bibitem{Achasov:2023gey}
M.~Achasov {\em et~al.}, {\em Front. Phys. (Beijing)} {\bf 19},   14701  (2024), \href{http://arxiv.org/abs/2303.15790}{{\ttfamily arXiv:2303.15790 [hep-ex]}}, \doi{10.1007/s11467-023-1333-z}.

\bibitem{Belle-II:2018jsg}
 Belle-II Collaboration (W.~Altmannshofer {\em et~al.}), {\em PTEP} {\bf 2019},   123C01  (2019), \href{http://arxiv.org/abs/1808.10567}{{\ttfamily arXiv:1808.10567 [hep-ex]}}, \doi{10.1093/ptep/ptz106}, [Erratum: PTEP 2020, 029201 (2020)].

\bibitem{CEPCStudyGroup:2018ghi}
 CEPC Study Group Collaboration (M.~Dong {\em et~al.}) (11 2018), \href{http://arxiv.org/abs/1811.10545}{{\ttfamily arXiv:1811.10545 [hep-ex]}}.

\bibitem{FCC:2018byv}
 FCC Collaboration (A.~Abada {\em et~al.}), {\em Eur. Phys. J. C} {\bf 79},   474  (2019), \doi{10.1140/epjc/s10052-019-6904-3}.

\end{thebibliography}
\end{document}